\documentclass[12pt,preprint,showpacs,preprintnumbers,amsmath,amssymb]{iopart}
\usepackage{amsfonts}
\usepackage{amsbsy}
\usepackage{graphicx}
\usepackage{xcolor}
\usepackage{verbatim}

\renewcommand{\vec}[1]{\mbox{\boldmath$\mathrm{#1}$}}
\newcommand{\tabrule}{\rule[-0.2em]{0em}{1.2em}}

\def\ind#1{{_{\mathrm{#1}}}}
\begin{document}

\title{Polarization and magnetization dynamics of a field-driven multiferroic structure}

\author{Alexander Sukhov$^1$, Chenglong Jia$^1$, Paul P. Horley$^2$ and Jamal Berakdar$^1$}

\address{$^1$Institut f\"ur Physik, Martin-Luther-Universit\"at Halle-Wittenberg, 06120 Halle, Germany \\ $^2$Centro de Investigati\'{o}n en Materiales Avanzados, S.C. (CIMAV), Chihuahua / Monterrey, 31109 Chihuahua, Mexico}

\begin{abstract}
We consider a multiferroic chain with
 a linear magnetoelectric coupling induced by the electrostatic screening at the ferroelectric/ferromagnet interface. We study theoretically  the dynamic  ferroelectric and magnetic
 response  to external magnetic and electric fields by utilizing an approach
 based on coupled Landau-Khalatnikov and  finite-temperature Landau-Lifshitz-Gilbert equations.
   Additionally, we compare with  Monte Carlo calculations. It is demonstrated that for material parameters corresponding to BaTiO$_3$/Fe
    the polarization and the magnetization are {controllable} by  external magnetic and electric fields respectively.
\end{abstract}

\maketitle
\section{Introduction}
\begin{center}
  \begin{figure}[htb]
   \centering
   \includegraphics[width=.3\textwidth]{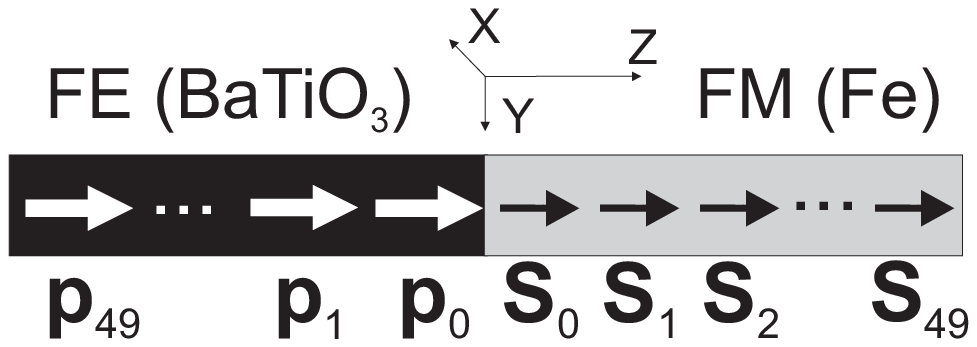}
   \caption{\label{fig_1} Alignment of electric dipoles $\vec{p}_j$ and magnetic moments $\vec{S}_i$ in the considered one-dimensional chain. The directions of $\vec{S}_i(t=0)\approx\{0,0,1\}$ and $\vec{p}_j(t=0)=\{0,0,1\}$ represent the initial configuration. The easy axis in the FM part is along the Z-direction.}
   \end{figure}
   \end{center}

Magnetic nanostructures are intensively researched\cite{HiOu010306}
due to their versatile use in technology.  \textit{Multiferroics}, i.e. systems
 that exhibit a coupled ferroelectric and ferromagnetic order, have received much attention
 recently\cite{Fi05,Sc06,Ra07}. A variety of potential applications of multiferroics rely on
   a possible control of magnetism (electric polarization) with electric (magnetic) fields due to the
   magnetoelectric coupling.
   Indeed, recent experiments  have demonstrated the existence of both effects\cite{Lo04,Ki03,alexe}.
Furthermore, a number of interesting  phenomena associated with
 the coupled polarization/magnetization dynamics at interfaces and in bulk have been
  reported such as the electrically controlled exchange bias\cite{Bo05}, the electrically controlled magnetocrystalline anisotropy\cite{We07} and  the influence of electric field  on the spin-dependent transport\cite{Ts06}.
Theoretically, recent Monte-Carlo calculations for a three-dimensional spinel lattice were performed
to study the magnetic-field induced polarization rotation\cite{Yao_epl,Yao_jap}.\\
In this work we investigate theoretically {and numerically} the field-driven dynamics of the electric polarization and the magnetization of a ferroelectric/ferromagnetic system that shows a magnetoelectric coupling at the interface.
For this purpose we consider a two-phase multiferroic chain consisting of {50 polarization sites and 50 localized magnetic moments}, as sketched in Fig.\ref{fig_1}. The ferromagnetic (FM) part of the chain is a normal metal (e.g., Fe), whereas the  ferroelectric (FE) part is BaTiO$_3$ (Fig. \ref{fig_1}). Recently this system has  been shown to exhibit a magnetoelectric coupling\cite{Du06,Fe08} and has been realized experimentally\cite{Sa07}.
The multiferroic coupling arises as a result of an accumulation of spin-polarized electrons or holes  at the FE-insulator/FM-metal interface when the  FE  is polarized\cite{Ro08}. At the metal/insulator interface, the screening of the polarization charge alters  the FE polarization orientation resulting in a linear change of the surface magnetization. The resulting magnetization in the FM structure decays exponentially away from the interface. Taking the FM material as an ideal metal {with} the screening length {of} around {\cite{Ki05scr}} $1\AA$, the exchange interaction between the additional surface magnetization and the FM part is thus limited to only the first site.
Switching to  {dimensionless} units, we introduce the reduced  polarization ($\vec{p}_j(t)=\vec{P}_j(t)/P\ind{S}$) and magnetic moment ($\vec{S}_i(t)=\vec{\mu}_i(t)/\mu\ind{S}$) vectors, where $P\ind{S}$ is the spontaneous polarization of (bulk) BaTiO$_3$  and $\mu\ind{S}$ is the  magnetic moment at saturation of (bulk) Fe.
\section{Theoretical formalism}
The total energy of the FE/FM-system {in a very general one-dimensional case} consists of three parts
\begin{equation}
\displaystyle F_{\Sigma}=F_{\mathrm{FE}}\, + F_{\mathrm{FM}} + E_{\mathrm{c}}.
\label{eq_1}
\end{equation}
The ferroelectric energy contribution reads
\begin{eqnarray}
\label{eq_2}
   F_{\mathrm{FE}}=a^3P\ind{S}\sum_{j=0}^{N_{\mathrm{FE}}{ -1}}\Big(&& \frac{\alpha_{\mathrm{FE}}P\ind{S}}{2}\,\vec{p}_j^2 + \frac{\beta_{\mathrm{FE}}P^3_{\mathrm{S}}}{4}\,\vec{p}_j^4 + \nonumber \\
   &&\frac{\kappa\ind{FE}_{ j} P\ind{S}}{2}\,\big(\vec{p}_{j+1} - \vec{p}_{j}\big)^2 - \vec{p}_j\cdot{ \vec{E}(t)}\Big),
\end{eqnarray}
whereas the ferromagnetic energy part is
\begin{equation}
\label{eq_3}
   F_{\mathrm{FM}}=\sum_{i=0}^{N_{\mathrm{FM}}{ -1}}\Big( -J_{ i} \,\vec{S}_i\cdot\vec{S}_{i+1}-D_{ i} \,(S_i^{\mathrm{z}})^2-{ \mu\ind{S}} \,\vec{S}_i\cdot \vec{B}(t)\Big).
\end{equation}
$\vec{E}(t)$ and $\vec{B}(t)$ are respectively external electric and magnetic fields.\\
Various pinning effects that may emerge in both the FE and the FM parts due to imperfections and expressed by $\kappa\ind{FE}_j$ and $J_i$, $D_i$ respectively, are not considered here, i.e. $\kappa\ind{FE}_j\equiv \kappa\ind{FE}$, $J_i \equiv J$, $D_i \equiv D$.\\
We consider that the linear FE/FM coupling appears as a result of the exchange interaction of the magnetization induced by the screening charge at the interface and the local magnetization in the ferromagnet (for details we refer to \cite{Cai09}) and can be written as
\begin{equation}
E_{\mathrm{c}}=\lambda \,P_{\mathrm{S}}\, \mu_{\mathrm{S}}\, \vec{p}_0\cdot \vec{S}_0.
\label{eq_4}
\end{equation}
The meaning of the quantities appearing in these equations is explained in Table \ref{tab_parameters}.\\
{Based on the parameters obtained from \textit{ab-initio} calculations for BaTiO$_3$/Fe-interface\cite{Du06,Du08}, we estimate the coupling constant as $\lambda=J\,a^2_{\mathrm{FM}}\,\alpha\ind{S}/(\varepsilon\ind{FE}\, \varepsilon_0 \mu_0 \mu^2_{\mathrm{S}})\approx 2\cdot 10^{-6}\,$s/F, where the surface ME-coupling constant is $\alpha\ind{S}=2\cdot 10^{-10}\, \mathrm{G\,cm}^2\mathrm{/V}$. We find this value is too low to obtain a sizable ME-response for the one-dimensional multiferroic interface. In what follows, we vary $\lambda$ and explore the dependence of the multiferroic
dynamics on it.}\\
 %
The polarization dynamics is governed by the Landau-Khalatnikov (LKh) equation\cite{Ma08,Ri98}, i.e.
\begin{equation}
\displaystyle \gamma_{\nu}P\ind{S} \frac{d \vec{p}_j}{dt} = \vec{H}_j^{\mathrm{FE}}= -\frac{1}{a^3 P\ind{S}}\frac{\delta F\ind{\Sigma}}{\delta \vec{p}_j},
\label{eq_5}
\end{equation}
where $\gamma_{\nu}$ is the viscosity constant (Table \ref{tab_parameters}) and $\vec{H}_j^{\mathrm{FE}}$ stands for the total external and internal fields acting on the local polarization. The magnetization dynamics obeys the Landau-Lifshitz-Gilbert\cite{La35Gi55} (LLG) equation of motion
\begin{equation}
\label{eq_6}
   \displaystyle \frac{d\vec{S}_i}{dt} = - \frac{\gamma}{1+\alpha^2\ind{FM}} \left[\vec{S}_i\times \vec{H}_i^{\mathrm{FM}}(t)\right]
   -\frac{\gamma \alpha\ind{FM}}{1+\alpha^2\ind{FM}} \left[\vec{S}_i\times\left[\vec{S}_i\times\vec{H}_i^{\mathrm{FM}}(t)\right]\right],
\end{equation}
where $\gamma$ is a gyromagnetic ratio (Table \ref{tab_parameters}) and $\alpha\ind{FM}$ is the Gilbert damping parameter. The total effective field acting on $\vec{S}_i$ is defined as a sum of deterministic and stochastic parts
$\displaystyle{\vec{H}_i^{\mathrm{FM}}(t) = -\frac{1}{{ \mu\ind{S}}}\frac{\delta F\ind{\Sigma}}{\delta \vec{S}_i} + \vec{\zeta}_i(t)}.$
The characteristics of the additive white noise associated with the thermal energy $k\ind{B}T$ are\cite{GaPa98}
   $   \displaystyle{\left<\zeta_{ik}(t)\right>=0} $ and $
      \displaystyle{\left<\zeta_{ik}(t)\zeta_{ml}(t + \Delta t)\right> = \frac{2\alpha\ind{FM}k\ind{B}T}{\mu\ind{S}\gamma} \delta_{im} \delta_{kl} \delta(\Delta t).}$
Here $i$ and $m$  index  the corresponding sites in the FM-material. $k$ and $l$ are the Cartesian components of $\zeta$ and $\Delta t$ is the time interval.
The coupled equations of motion (\ref{eq_5}) and (\ref{eq_6}) are solved numerically in reduced units, re-normalizing the energy (\ref{eq_1}) over doubled anisotropy strength $D$. Thus, the dimensionless time in  both equations is $\tau=\omega\ind{A}\,t=\gamma\, B\ind{A}\,t=\gamma \, 2D\, t/\mu\ind{S}$ and the reduced effective fields are $\vec{h}_i^{\mathrm{FM}}(\tau)=\vec{H}_i^{\mathrm{FM}}(\tau)/B\ind{A}$, $\vec{h}_j^{\mathrm{FE}}=\vec{H}_j^{\mathrm{FE}}/(\gamma \gamma\ind{\nu}P\ind{S}B\ind{A})$.\\
To endorse our  results we conducted furthermore  kinetic Monte Carlo (MC) simulations for the description of the dynamics of a FE/FM chain subjected to  external magnetic and electric  fields. The electric dipoles and the magnetic moments are understood as three-dimensional classical unit vectors, which are randomly updated with the standard Metropolis algorithm\cite{HiNo98}. The period of the external field is chosen to be 600 MC steps per site.\\
In the following the system is described via the reduced total polarization $\vec{p}\ind{\Sigma}(t)$
\begin{equation}
\displaystyle \vec{p}\ind{\Sigma}(t)=\frac{1}{N\ind{FE}}\sum_{j=0}^{N\ind{FE}{ -1}} \vec{p}_j(t),
\label{eq_9}
\end{equation}
and the reduced net magnetization $\vec{S}_{\Sigma}(t)$
\begin{equation}
\displaystyle \vec{S}\ind{\Sigma}(t)=\frac{1}{N\ind{FM}}\sum_{i=0}^{N\ind{FM}{ -1}} \vec{S}_i(t).
\label{eq_10}
\end{equation}
\section{Results of numerical simulations}
\begin{center}
  \begin{figure}[htb]
   \centering
   \includegraphics[width=.6\textwidth,angle=-90]{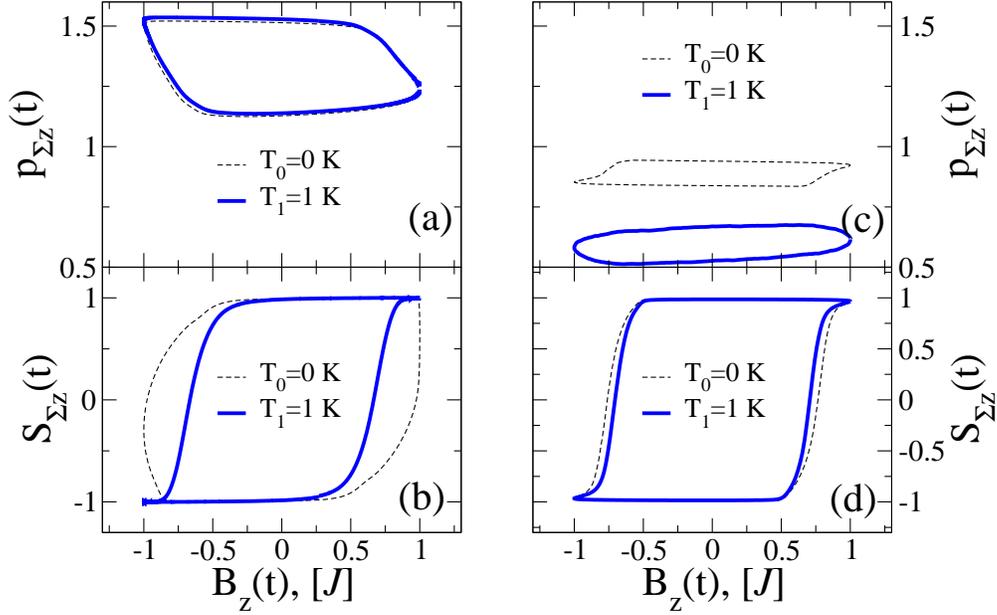}
   \caption{\label{fig_2} The reduced total polarization/magnetization response to external magnetic field of the form $B\ind{z}(t)=B\ind{0z}\cos \omega t$. The loops a) and b) are obtained by using the LKh and the LLG equations; c) and d) are calculated using the MC method. In the both methods parameters are chosen such that $E\ind{c}\approx\mu\ind{S}B\ind{z}\approx J$, i.e.: $\lambda=240$~s/F, $B\ind{0z}=6.65 B\ind{A}$, $\omega=3.61\cdot 10^{11}\,\,\mathrm{s}^{-1}$, $E\ind{0z}=0$~V/m, $\alpha=0.5$. 20 first periods ($1/\omega$) are omitted; the hysteresis curves are averaged over 100 (a, b) and 200 (c, d) subsequent periods.}
   \end{figure}
\end{center}

\begin{center}
  \begin{figure}[htb]
   \centering
   \includegraphics[width=.6\textwidth,angle=-90]{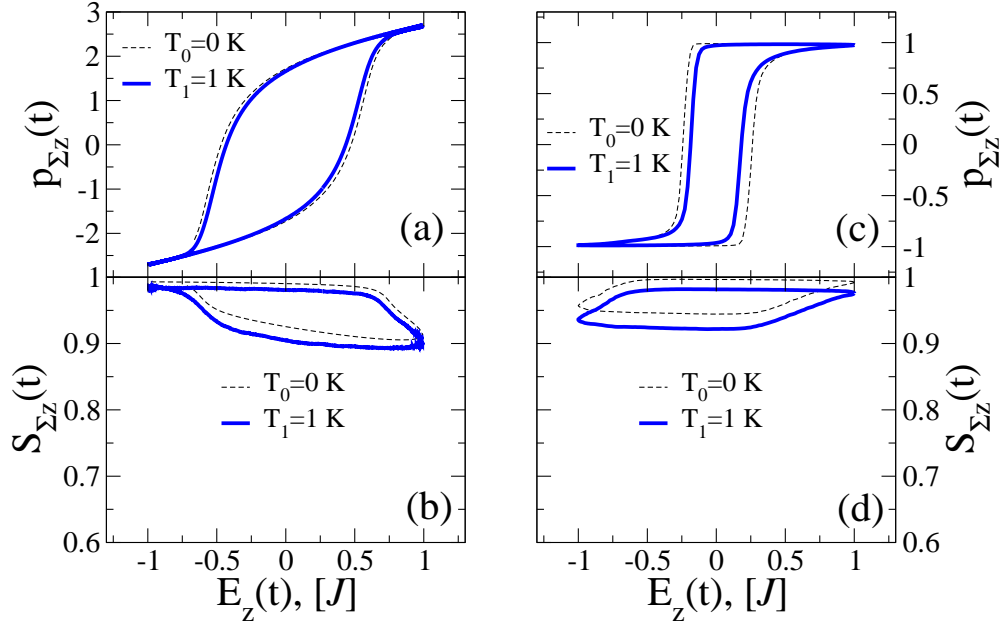}
   \caption{\label{fig_3} Hysteresis loops of the total reduced polarization/magnetization as a function of external electric field of the form $E\ind{z}(t)=E\ind{0z}\cos \omega t$. The curves a) and b) are obtained by using the LKh and the LLG equations; c) and d) are calculated using the MC method. Parameters are chosen such that $E\ind{c}\approx a^3P\ind{S}E\ind{z}\approx J$, i.e.: $\lambda=240$~s/F, $B\ind{0z}=0$~T, $E\ind{0z}=4.07\cdot 10^7$~V/m, $\omega=3.61\cdot 10^{11}\,\,\mathrm{s}^{-1}$, $\alpha=0.5$. 20 first periods ($1/\omega$) are omitted; the hysteresis loops are averaged over 100 (a, b) and 200 (c, d) subsequent periods.}
   \end{figure}
\end{center}

To demonstrate the response of the FE/FM-chain to external fields using the LKh and the LLG equations we used a
damping parameter $\alpha\ind{FM}=0.5$, which is significantly higher than the experimental value\cite{Oo06} ($\alpha$(Fe)=0.002), in order to achieve a faster relaxation of the net magnetization for both B- and E-drivings.
It was assured in our calculations that the FE-subsystem is far from its phase transition temperature at $T=10$~K and the FM-subsystem remains non-superparamagnetic on the whole time scale of consideration.\\
Fig. \ref{fig_2} shows the hysteresis loops in the presence of a harmonic external magnetic field $B\ind{z}(t)=B\ind{0z}\cos \omega t$. The amplitude of the field $B\ind{0z}$ is chosen to be comparable with the exchange interaction energy $J$ as well as the coupling energy $E\ind{c}$. The period of the external magnetic field is chosen to exceed the field-free precessional period ($2\pi/\omega\approx 5T^{\mathrm{prec}}$) of the LLG. Irrespective of the temperature and the calculational method this field is capable of switching the magnetization of the FM-chain (Fig. \ref{fig_2}b, d). The FE-polarization indirectly driven by the external magnetic field is not completely switched according to  both methods (Fig. \ref{fig_2}a, c). The role of thermal
fluctuations on the FM part only (cf. eq. (\ref{eq_5})) is exposed by the p(B)-behavior shown in Fig. \ref{fig_2}a, whereas  the MC method accounts for temperature  effects also on the polarization  (Fig. \ref{fig_2}c).  {As a result, the $p(B)$-hysteresis (Fig. \ref{fig_2}c) shows a clear temperature dependence, which becomes especially pronounced for the one-dimensional chain in which thermal fluctuations degrade the polarization/magnetization ordering more intensively than for the case of a two-dimensional system.}\\
The hysteresis loops for the external electric field of the form $E\ind{z}(t)=E\ind{0z}\cos \omega t$ are presented in Fig. \ref{fig_3}. The energy of the applied electric field is comparable with the coupling energy. As a result, the total polarization can be completely switched (Fig. \ref{fig_3}a, c). As inferred from Fig. \ref{fig_3}b, d the net magnetization is not fully switched. Only several first spin sites follow the electric field due to the coupling at the interface.
\section{Discussion}
{In real experiments several additional effects may affect the polarization and the magnetization dynamics. Below we estimate how strong these effects might be for the multiferroic chain.}
\subsection{Effect of depolarizing fields}
In the general case {the field $\vec{E}(t)$ entering equation (\ref{eq_2}) is an effective
 field that consists of the applied electric field, e.g. $E\ind{z}(t)\vec{e}\ind{z}$, and the internal depolarizing field $\vec{E}\ind{DF}$ created by the screening charges (SC) at the interface.\\
 Generally, the depolarizing field  may well be sizable and affects thus the dynamics
 \cite{Me73,Kim05,Per07}. Here we estimate the strength of $\vec{E}\ind{DF}$ by introducing a one-dimensional SC at the interface $Q\ind{SC}=P\ind{S}a^2$ (Table \ref{tab_parameters})}. As a result, the electric field induced by the SC is opposite to the local polarization (Fig. \ref{fig_1}) and can be written as $\vec{E}\ind{DF}= - Q\ind{SC}/(4\pi \varepsilon_0 \varepsilon \ind{FE}a^2 n_j^2) \vec{e}\ind{z}$, where $\varepsilon_0=8.85 \cdot 10^{-12}$~A$\cdot$s/(V$\cdot$m) is the permittivity of free space, $\varepsilon\ind{FE}\approx 2000$ is the dielectric constant in barium titanate and $n_j$ is the index numbering the polarization sites starting from the interface, e.g. $n_{j=0}=1$. Thus, the strength of the depolarizing field calculated for $n_{j=0}=1$ and upon the other parameters is $E\ind{DF}\approx 2\cdot 10^6$~V/m, which is at least one order of magnitude smaller than the amplitudes of the applied electric field ($\approx 4\cdot 10^7$~V/m). Keeping in mind that $E\ind{DF}$ decays in the FE away from the interface we can neglect the depolarizing field.
\subsection{Effect of induced electric and magnetic fields}
{According to the Maxwell's equations an oscillating magnetic field induces an oscillating electric field, and an alternating voltage produces an oscillating magnetic field. The situation becomes especially important for the second case, since even small induced magnetic fields aligned perpendicularly to the inducing field and hence to the initial state of the magnetization can sufficiently assist the switching at appropriate frequencies\cite{SuWa_prb_73,SuWa_prb_74}.\\
From the Faraday's law of the Maxwell's equations $\nabla \times \vec{E} = - \mu\ind{FE}\, \mu_0 \frac{\partial \vec{H}}{\partial t}$ and for the given applied magnetic field $B\ind{z}(t)=B\ind{0z}\cos \omega t$ the induced electric field acting on the FE-polarization is oriented perpendicularly to the inducing field ($XY$-plane, Fig. \ref{fig_1}). Its amplitude for relative magnetic permittivity in BaTiO$_3$\cite{Wa_jac_454} $\mu\ind{FE}\approx 1$ and $\mu\ind{FM}\approx 5000$ scales as $E^{\mathrm{ind}}_{\mathrm{0}}=a\ind{FE}\, \mu\ind{FE}/(\mu\ind{FM})\,B\ind{0z}\,\omega\,\approx 2\,\mathrm{V/m}$.\\
Likewise, when an external electric field $E\ind{z}(t)=E\ind{0z}\cos \omega t$ is applied, according to the Ampere's law of the Maxwell's equations $\nabla \times \vec{B} = \mu\ind{FM}\,\mu_0\,\varepsilon\ind{FM}\,\varepsilon_0\, \frac{\partial \vec{E}}{\partial t}$, the direction of the induced field is perpendicular to the inducing field. The amplitude of the induced magnetic field in iron for $a\ind{FM}=0.28\cdot 10^{-9}\,\mathrm{m}$ and $\varepsilon\ind{FM}\approx 1$ (since iron is assumed as an ideal metal) is $B^{\mathrm{ind}}_{\mathrm{0}} = a\ind{FM}\, \mu\ind{FM}\, \mu_0\, \varepsilon \ind{FM}\, \varepsilon_0\, E\ind{0z}\,\omega\,\approx 2\cdot 10^{-3}\, \mathrm{T}$.\\
Our calculations with the estimated (and even larger) amplitudes of the induced electric field show no influence on the Z-projection of the FE-polarization. This is a consequence of the uncoupled nature for the projections of the FE-polarization (cf. equation (\ref{eq_5})). Both numerical methods give a slightly enhanced (less than 1\%) ME-response in the presence of the induced magnetic field. Therefore, the effect of the induced electric and magnetic fields can be deemed irrelevant  for the considered multiferroic chain and for the chosen range of frequencies.}
\subsection{Frequency dependence of the magnetoelectric response}
\begin{center}
  \begin{figure}[htb]
   \centering
   \includegraphics[width=.6\textwidth,angle=-90]{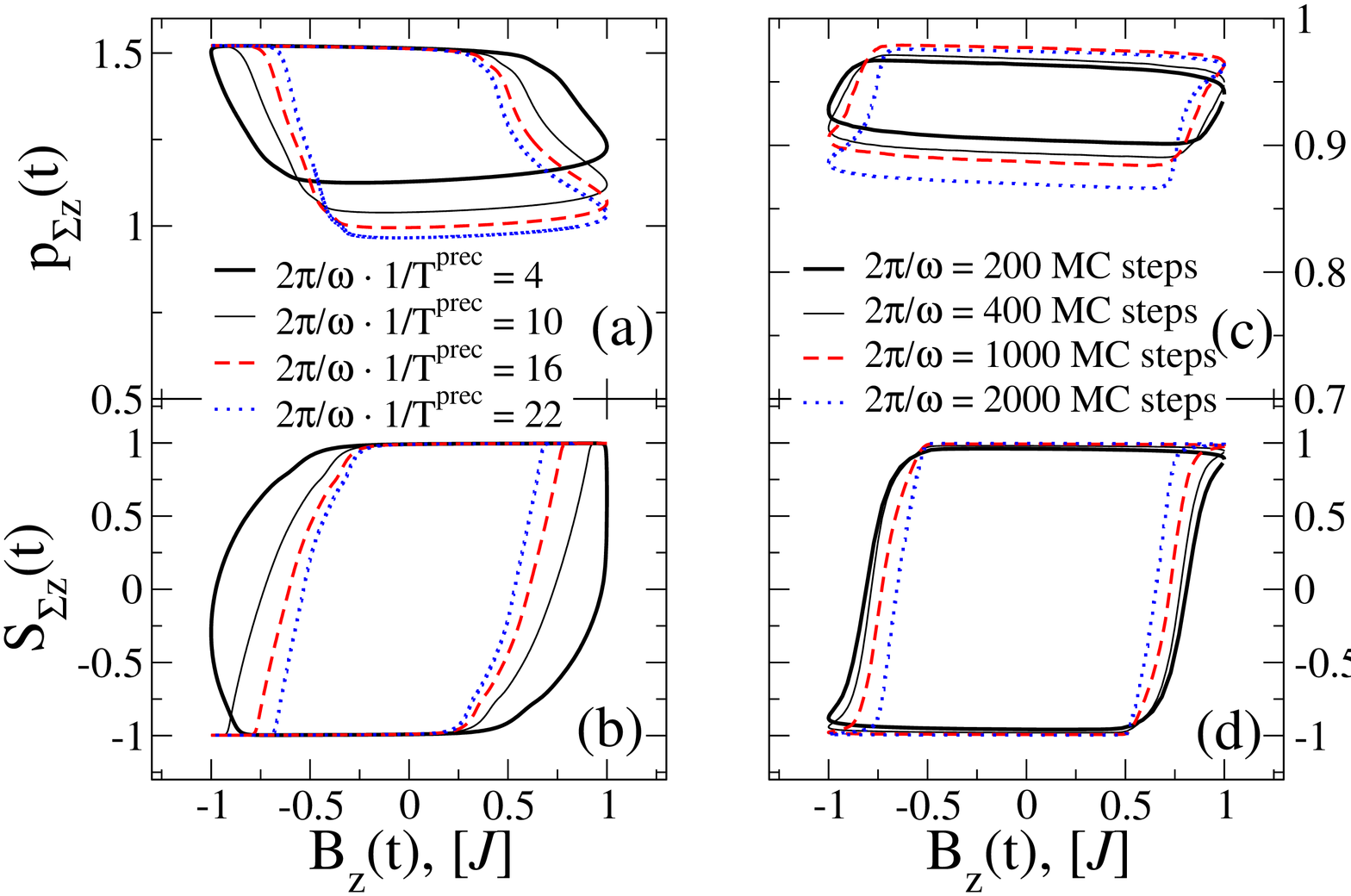}
   \caption{\label{fig_4} Response of the one-dimensional multiferroic structure to the time-dependent magnetic field $B\ind{z}(t)=B\ind{0z}\cos \omega t$ plotted for various frequencies $\omega$. The field-free precessional time in the FM part is $T^{\mathrm{prec}}_{\mathrm{FM}}=T^{\mathrm{prec}}\approx 4$~ps. The curves a) and b) are obtained by using the LKh and the LLG equations; c) and d) are calculated using the MC method. Parameters are $T_0=0$~K, $\alpha=0.5$, $B\ind{0z}=6.65 B\ind{A}$ and $\lambda=240$~s/F. Several first periods are omitted.}
   \end{figure}
\end{center}

\begin{center}
  \begin{figure}[htb]
   \centering
   \includegraphics[width=.6\textwidth,angle=-90]{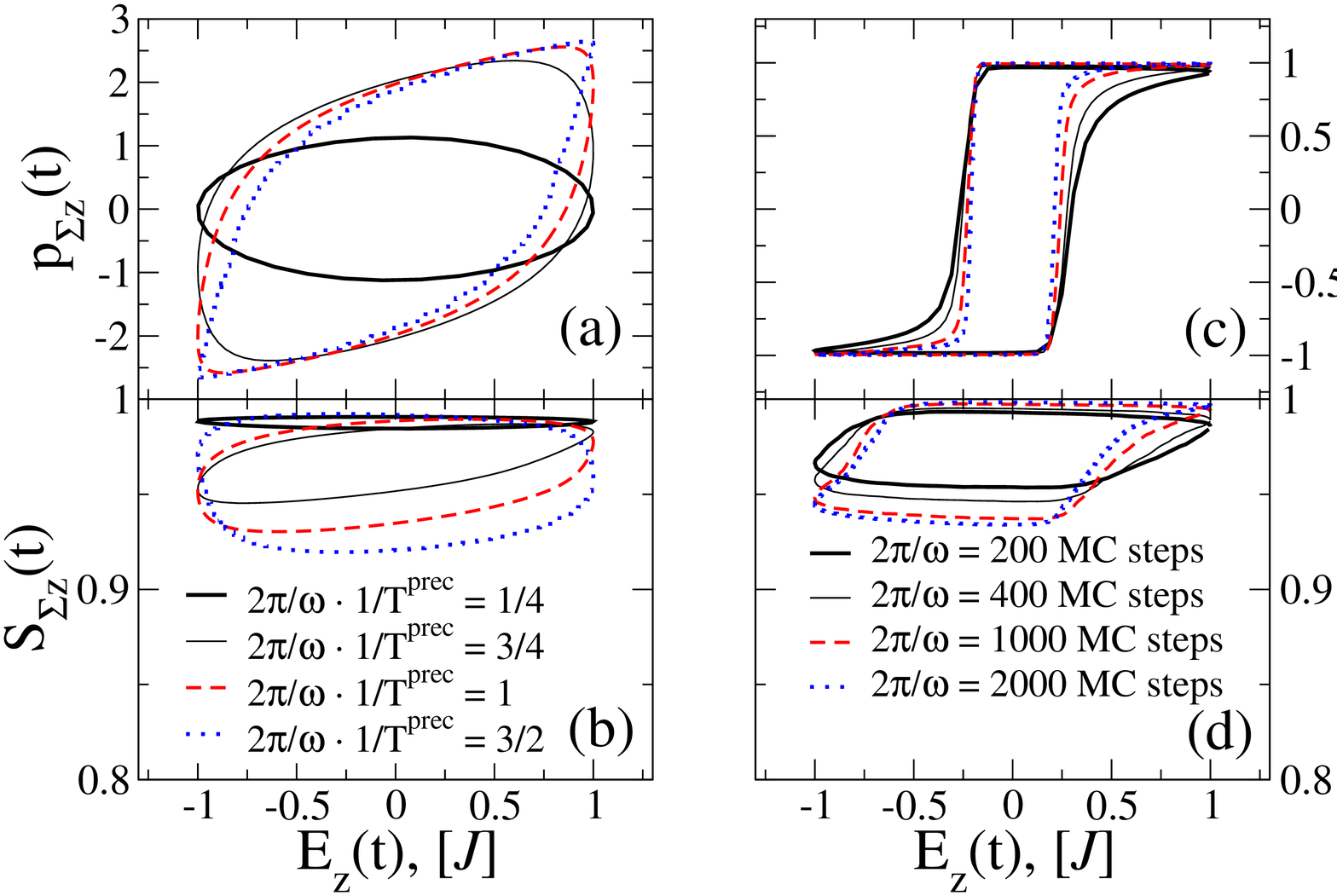}
   \caption{\label{fig_5} Response of the one-dimensional multiferroic structure to the time-dependent electric field $E\ind{z}(t)=E\ind{0z}\cos \omega t$ plotted for various frequencies $\omega$. The field-free precessional time in the FM part is $T^{\mathrm{prec}}_{\mathrm{FM}}=T^{\mathrm{prec}}\approx 4$~ps. The curves a) and b) are obtained by using the LKh and the LLG equations; c) and d) are calculated using the MC method. Parameters are $T_0=0$~K, $\alpha=0.5$, $E\ind{0z}=4.07\cdot 10^7$~V/m and $\lambda=240$~s/F. Several first periods are omitted.}
   \end{figure}
\end{center}

{A variation of the frequency $\omega$ of the external electric and magnetic fields can also affect the ME-response of the multiferroic chain.\\
Fig. \ref{fig_4}  demonstrates the response of the multiferroic interface to an external magnetic field. The periods of the external magnetic field should be compared with a characteristic field-free precessional time $T^{\mathrm{prec}}_{\mathrm{FM}}\approx 4$~ps which is valid for bulk iron. As one expects, magnetic fields with longer periods favor better saturation of the magnetization (Fig. \ref{fig_4}b). The response of the total electric polarization to the external magnetic field becomes enhanced with increasing period of B-field. This is confirmed by both numerical methods (Fig. \ref{fig_4}a, c).\\
The multiferroic response to an external electric field is shown in Fig. \ref{fig_5} for which the situation of very short electric fields (less or around $T^{\mathrm{prec}}_{\mathrm{FM}}$) is addressed. The magnetization does not relax quick enough resulting in the form of a hysteresis which is similar to the ferroelectric one. Additionally, we obtain an increase of the net magnetization response (Fig. \ref{fig_5}b). This feature can also be observed using the MC-method (Fig. \ref{fig_5}d).}
\section{Summary}
The main result obtained using two independent methods - the direct solution of the LKh and the LLG equations (\ref{eq_5}, \ref{eq_6}) and the kinetic MC method - is that due to the  coupling at the interface of FE/FM  the ferromagnetic subsystem responds to an external electric field and the ferroelectric subsystem responds to an external magnetic field. A use  of  both methods allowed a comparison of dynamical and statistical approaches for studying coupling phenomena at the FE/FM interface. Additionally, the LKh/LLG equations provide an insight into the real time temporal behavior, while MC approach is very useful to inspect the temperature influence on both sides of the interface.\\
This research is supported by the research projects DFG SFB762 (Germany) and FONCICYT 94682 (Mexico).\\

\begin{table}[htb]
\caption{Parameters used in the numerical calculations.}
\centering
\vspace{2.ex}
\begin{tabular*}{0.6\textwidth}{@{\extracolsep{\fill}}lr}
 \hline
 \multicolumn{2}{c}{ FE-material (BaTiO$_3$)}\\
 \hline
 \hline
 Number of sites $N\ind{FE}$ \tabrule& 50 \\
 \hline
 Polarization\cite{Hl06} $P_{\mathrm{S}}$, [C/m$^2$]\tabrule& 0.265  \\
 \hline
 Initial state $\vec{p}_j$(t=0), [P$_{\mathrm{S}}$]\tabrule& \{0.0,0.0,1.0\}  \\
 \hline
 Constant\cite{Ma08} $\gamma_{\nu}$, [Vms/C]\tabrule& 2.5$\cdot10^{-5}$\\
 \hline
 Constant\cite{Hl06} $\alpha_{\mathrm{FE}}$, [Vm/C]\tabrule& -2.77$\cdot 10^{7}$\\
 \hline
 Constant\cite{Hl06} $\beta_{\mathrm{FE}}$, [Vm$^5$/C$^3$]\tabrule& 1.70$\cdot 10^{8}$\\
 \hline
 FE-interaction $\kappa_{\mathrm{FE}}$, [Vm/C]\tabrule& 1.0$\cdot 10^8$\\
 \hline
 Lattice constant\cite{Jo57} $a$, [m] \tabrule& 0.4$\cdot$ 10$^{-9}$ \\
 \hline
 FE/FM-coupling $\lambda$, [s/F] \tabrule& $\mathrm{parameter}$ \\
 \hline
 \multicolumn{2}{c}{ FM-material (Fe)}\\
 \hline
 \hline
 Number of sites $N\ind{FM}$ \tabrule& 50 \\
 \hline
 Gyromagn. ratio $\gamma$, [(Ts)$^{-1}$] \tabrule& $1.76\cdot10^{11}~$\\
 \hline
 Moment per site\cite{Vi74} $\mu\ind{S}$, [$\mu_{\mathrm{B}}$]\tabrule& 2.2\\
 \hline
 Initial state $\vec{S}_i$(t=0), [$\mu\ind{S}$]\tabrule& \{0.14,0.14,0.98\}  \\
 \hline
 Anisotropy strength\cite{Ki05} $D$, [J]\tabrule& 1.0$\cdot 10^{-22}$\\
 \hline
 Exchange strength\cite{Ki05} $J$, [J]\tabrule& 1.33$\cdot 10^{-21}$\\
 \hline
 Damping $\alpha_{\mathrm{FM}}$\tabrule& $\mathrm{parameter}$\\
 \hline
\end{tabular*}
\label{tab_parameters}
\end{table}
\end{document}